\documentclass{aa}

\usepackage{natbib}
\usepackage{graphics}
\usepackage[mathcal]{eucal}
\usepackage{amssymb} 

\newcommand{\ds}{\displaystyle}
\newcommand{\fig}[3]{
      \begin{figure}[tbp]
	\resizebox{\hsize}{!}{\includegraphics  {#1}}
	\caption{#2}
	\label{#3}
        \end{figure} }

\newcommand{\inv} {\frac {1}}

\newcommand{\derivp} [2] {\frac {\partial #1 } {\partial #2} }

\newcommand{\eqn} [1] {
\begin{equation} 
#1 
\end{equation}}

\newcommand{\eqna} [1] {
\begin{eqnarray} 
#1 
\end{eqnarray}}

\begin{document}


\title{Excitation of stellar p-modes  by turbulent convection : }
\subtitle {2. The Sun}

\author{Samadi R. \inst{1} \and   Goupil M.-J.\inst{2} \and 
Lebreton Y. \inst{2} }

\institute{
Observatoire de Paris. DESPA. CNRS UMR 8632. 92195 Meudon. France \and
Observatoire de Paris. DASGAL. CNRS UMR 8633. 92195 Meudon. France 
}
\offprints{R. Samadi}
\mail{reza.samadi@obspm.fr}

\date{Received 17 October 2000 / Accepted 19 January 2001}

\abstract{Acoustic power 
 and oscillation amplitudes of radial oscillations 
computed for a solar model are compared 
with solar seismic observations. The oscillations 
are assumed to be stochastically
 excited by turbulence.
The numerical computations are based upon a 
  theoretical formulation of the power going into solar-like 
oscillation modes, as proposed 
by   \citet{Samadi00I} in a companion paper. 
This formulation allows us to investigate
several  assumptions concerning properties of the stellar turbulence. 
We find  that the entropy source plays a  dominant role in the stochastic
 excitation compared with the Reynold stress source, in agreement 
with  \citet{GMK94}.
We consider several turbulent kinetic energy spectra suggested 
by different observations of the solar granulation.
Differences between turbulent spectra manifest in
large differences  in the computed oscillation powers 
at high oscillation frequency.
Two free parameters which  are introduced in the description of the turbulence 
enter  the expression for the acoustic power. These parameters 
are  adjusted  in order to fit to the solar 
observations of the surface velocity oscillations.
The best fit is obtained with the kinetic
 energy spectrum deduced from the observations
 of the solar granulation by \citet{Nesis93}; 
the corresponding adjusted parameters are found to 
be compatible with the theoretical upper limit which can be set 
on these parameters. The  adopted theoretical approach improves the 
agreement between solar seismic observations and  
numerical results.
\keywords{convection - turbulence - Stars : oscillations - Sun: oscillations}
}

\maketitle

\section{Introduction}

It is currently believed that 
oscillations of solar-like stars can be 
 excited by the turbulent convection of their outer layers. 
The physical description of this stochastic excitation
which yields the power 
 transferred into the oscillations by the
 turbulence 
is not well known : 
the turbulent Reynolds stress 
was  first identified  \cite[hereafter GK]{GK77} 
as the main source term. 
In the last decade, turbulent 
entropy fluctuations had been proposed 
as a possible additional source term \citep{Stein91,Balmforth92c,GMK94}.

In Paper~I \citep{Samadi00I}  
we reconsider the excitation  of 
stellar p-modes  by turbulent convection.  
A new formulation following 
 the GK
approach and  taking into account 
both Reynolds stress and entropy fluctuations contribution  
is proposed. In contrast with previous works, the present
  formulation  is valid for any turbulent
 time spectrum and turbulent energy spectrum. 
The entropy source term was considered 
as the main source term by \citet[hereafter GMK]{GMK94} 
while  \citet[hereafter B92]{Balmforth92c} 
considered the corresponding contribution as negligible. 
However it was shown in Paper~I that the entropy 
source term considered by GMK and B92 vanishes  
and a  non-linear term involving the turbulent 
velocity field and the turbulent entropy 
represents the actual entropy contribution. 
Inherent to an empirical description of the stellar turbulence,
our theoretical model for the acoustic power involves  two free parameters, 
the  parameter $\lambda$ which is related to 
the rather  arbritary definition of eddy correlation 
time and the  parameter  $\beta$ 
which relates the mixing length 
to the largest wavenumber in the inertial range.
 
Several space seismic experiments  are 
currently  planed~: COROT \citep{Baglin98}, MONS \citep{Kjeldsen98} and MOST \citep{Matthews98}.
These projects will provide very accurate  seismic data
such as oscillation amplitudes
and damping rates for  a large set 
of stellar targets. The oscillation powers deduced from these
observations will provide valuable constraints on the current
physical understanding  of the excitation
by turbulent convection. 
In the meantime, we need 
 to validate
  and constrain our theoretical  approach with current 
solar observations.  This is the primary goal of this paper.
We first compute the oscillation power for a solar model (Sect.~2)
using conventional assumptions for the turbulent 
components and compare it  with observations.
In  Sect.~3 we investigate 
the influence of   the atmosphere  and of different turbulent
components such as the time spectrum, the turbulent spectrum and
the free parameters.

In Sect.~4, we discuss on the best treatment 
for the turbulent components  and then calibrate  the 
corresponding formulation on solar observations i.e.
determine  the free parameters
in view of applications to other solar-like
stars.

\section{Oscillation power}

\subsection{Theoretical formulation and  observations}

The power going into each mode of frequency $\omega_0$ is 
expressed as (see Paper~I)
\eqn{
P_{\omega_0} = \Gamma_{\omega_0} \; E_{\omega_0} 
\label{eqn:P_omega0}
} 
where $  \Gamma_{\omega_0} = 2 \eta$ is the linewidth of the mode,  $\eta$ is the  damping rate and $E_{\omega_0} $ is the averaged mode energy given by 
\eqn{
E_{\omega_0} =     \inv {2} \langle \mid A \mid ^2 \rangle  \; I {\omega_0}^2 ,
\label{eqn:E_omega0}
} 
where
\eqn{
I \equiv   \int_0^{M} dm \,  \vec \xi^*  . \vec \xi
}
is the mode inertia and $\langle \mid A \mid ^2 \rangle $ is the mean-square 
amplitude resulting from a balance between the excitation 
by the turbulent convection and damping processes. 
This amplitude  can be expressed in a schematical form as 
\eqn{
\left < A^2 \right > \propto\eta^{-1}\int_{0}^{M}dm \; \rho_0 \; w^4 \;\left
(\derivp { \xi_r} {r} \right )^2  \;   \mathcal{S}(\omega,m)
\label{eqn:A2}
} 
where $\displaystyle{\xi_r }$ is the radial 
displacement eigenfunction, $\rho_0$ the density, 
$w$  the vertical rms velocity of the convective
 elements and $\mathcal{S}$ the turbulent  source 
function 
including  both the Reynolds and the entropy fluctuation contributions.
Detailed expressions for $\left < A^2 \right >$ and $\mathcal{S}$ are given 
in Paper~I. The source function   involves the kinetic 
energy spectrum $E(k)$, the entropy energy spectrum $E_s(k)$ 
and the time spectrum $\chi_k(\omega)$. 
The turbulent spectra in $\mathcal{S}$ are integrated 
over all eddy wavenumbers $k$,  and $\mathcal{S}$ 
is in turn integrated in Eq.(\ref{eqn:A2}) over the stellar mass $M$.

The time spectrum  $\chi_k(\omega)$   
measures the temporal coherence of eddies with given $k$. 
The linewidth of this function is related 
to the time correlation $\tau_k$ associated 
with eddies with wavenumber $k$. $\tau_k$ is related to $k$ and the 
eddy velocity $u_k$ by (see Eq. 97 of Paper~I)
\eqn{
\tau_k= \lambda \, (k u_k)^{-1}
}
where the free  parameter $\lambda$ is introduced because 
the definition of the eddy correlation time $\tau_k$ and
 the evaluation  of the eddy velocity  are somewhat arbitrary. 
The eddy velocity $u_k$ is given by Eq.(98) of Paper~I. 
Several different time spectra are investigated later in this paper. 
In addition, we evaluate effect of changes in 
the parameter $\lambda$, which is usually set  to one, for instance in  B92.

The kinetic  energy spectrum $E(k)$ describes the energy carried 
by the turbulent eddies.
In a highly turbulent medium, 
such as in the outer stellar convection zone, it is expected 
from  turbulence theory that  such a spectrum exhibits 
an inertial range where the energy carried 
by the eddies with wavenumber $k$ decreases 
as  $k^{-5/3}$ in the Oboukhov-Kolmogorov theory. 
The wavenumber $k_0$ at which the turbulent cascade 
begins is not well determined. This quantity has 
been related in Paper~I to the mixing length $\Lambda$ as
\eqn{
k_0 =  \frac{2 \pi} {\beta \Lambda} \; ,
\label{eqn:k_0}
}
where  $\beta$  is a free parameter which 
has been introduced in order to gauge the arbitrary nature of such a relation. 
The injection region  lies in the $k \leq k_0$ range. 
Here we investigate several turbulent kinetic energy spectra 
exhibiting different forms in the injection region. 

The mean-square amplitude is inversely proportional to the damping rate 
\citep[e.g.][]{Balmforth92c} and consequently the acoustic noise generation rate 
(or acoustic power) $P_{\omega_0}$ is independent of $\eta$. Thus comparing
the estimated acoustic noise generation rates 
with observations avoids the additional
uncertainties in the modeling of the damping rates (GMK) but requires 
accurate linewidth measurements and model computations.

The  mean-square surface velocity of a mode is 
given by the relation (see Paper~I)
\eqn{
v_s^2(\omega_0)  =  \omega_0^2 \xi^2(r_s)  \; \inv{2} \langle \mid A \mid ^2 \rangle  =  \xi^2(r_s) \frac{P_{\omega_0}}{2 \eta I} 
\label{eqn:vs}
}  
where $r_s$ is the radius  of the layer seen by the instrument. 
Observations of $v_s$ performed by  \citet{Libbrecht88} 
are  plotted in Fig. \ref{fig:Vk} with
 accompanying error bars. The surface velocity amplitude $v_s$
steeply increases with frequency  at low frequencies, 
presents a maximum value around $\nu \simeq 3.2$~mHz and undergoes
a steep decline at high frequency. 
In Sect.~4 we  make use of these three characteristics 
to constrain the turbulent ingredients and the free parameters entering the
theory of the excitation process.

\subsection{Oscillation power in the solar case}

The solar model we consider has been calculated with the CESAM code \citep{Morel97} and appropriate input physics, described in details in  \citet{Lebreton99}. In particular, convection is described according to the classical mixing-length theory \cite[hereafter MLT]{Bohm58} with a mixing-length $\Lambda= \alpha_c \, H_p$, where $H_p$ is the pressure scale height and  $\alpha_c$ is the mixing-length parameter. 
The external boundary conditions are obtained through a $T(\tau)$ law derived from ATLAS9 model atmospheres \citep{Kurucz91}.
The solar metal to hydrogen mass ratio $\left (Z/X \right )_\odot=0.0245$ comes from  \citet{Grevesse93}.
The calibration of the solar model, in luminosity ($L_\odot= 3.846 \, 10^{33}$~erg/s) and radius ($R_\odot=6.9599 \, 10^{10}$~cm) for an age of $4.6$~Gyr, fixes the initial helium content $Y=0.2682$, metallicity $Z=0.0175$ and the MLT parameter  $\alpha_c=1.785$.
 
 The estimated cut-off frequency is about $\nu_c \simeq 5.32$~mHz. 
The oscillation eigenfunctions are obtained with the adiabatic 
 pulsation  FILOU code of \citet{Tran95}. 

We evaluate  the root mean-square surface 
velocity $v_s$ in Eq.(\ref{eqn:vs}) 
where the oscillation power $P_{\omega_0}$ 
is computed according to 
Eqs(\ref{eqn:P_omega0} , \ref{eqn:E_omega0} , \ref{eqn:A2}) 
and $\eta$ is given by solar observations of 
  \citet{Libbrecht88}. These solar
 observations sample the solar atmosphere at 
the  optical depth  $\tau_{5000} \simeq 0.05$. 
In our solar model this optical depth is localized
 approximatively at 140 km above the solar 
photosphere and $v_s$ is thus computed at this depth.

Fig. \ref{fig:Vk} shows  the resulting 
values of $v_s$ assuming the Kolmogorov spectrum 
as defined in Eq.(\ref{eqn:kolmo}). 

As in GMK and B92, 
it is assumed that the entropy spectrum exhibits the 
same behavior than the kinetic energy spectrum.  
Thus the entropy energy spectrum $ E_s(k)$  
considered in the computation is at this stage 
simply proportional to the kinetic energy spectrum $E(k)$ 
(Eq. 107 of Paper~I).
The parameter $\lambda$ is first set to one 
and the parameter $\beta$ is set to $1.9$,
according to the theoretical estimate obtained in Paper~I. 
The anisotropic factor $\Phi$ involved in the 
formulation may be considered as a free parameter. 
However in order to be consistent with Bohm-Vitense's 
MLT considered here, we set $\Phi=2$ in the present work.
$\Phi$ affects the $\nu$ dependence of the power in a similar way as $\lambda$ and $\beta$. Changing its value is therefore equivalent to change values of $\lambda$ and $\beta$,  which will be adjusted later (section \ref{sec:Solar calibration}).

At low frequencies, $v_s$ increases because 
the inertia of the modes decreases 
with frequency and 
because the derivatives of the displacement eigenfunctions increase 
with frequency (see GMK).
At high frequency,  $v_s$  decreases because  
the depth of the excitation  region 
and the efficiency  of the excitation 
decrease with frequency.
This  can be  explained  as follows: 
let $\tau_k$ be the characteristic lifetime of
 an eddy of wavenumber $k$. The major contribution to mode 
excitation comes from the eddies
 with $ \tau_k \omega_0 \lesssim 1$  \cite[Paper~I]{Goldreich90}. 
When $\omega_0$ increases, the region where $ \tau_k \omega_0 \lesssim 1$ 
becomes thinner and is confined at the top of the convection zone.
Futhermore, contributive eddies have smaller size and are less energetic.
Thus $P_{\omega_0}$ becomes smaller as $\omega_0$ increases.

Including only the Reynolds stress leads to amplitudes much 
smaller than those observed. 
The contribution of  the entropy source term 
is found to be larger than the Reynolds stress contribution 
and increases the amplitudes. 
This result is  in agreement with the conclusion of GMK 
and with the simulations of  \citet{Stein91}.
However, some discrepancy 
still remains between the 
theoretical estimates and the observations. 
In particular the  theoretical absolute amplitudes remain 
much smaller than the observed ones 
 and large differences are observed at high frequency 
in  terms of frequency dependence between the  
amplitudes computed using the Kolmogorov spectrum and the observations.
Also, the maximum amplitude peaks  at a frequency which 
is smaller by about 0.1 mHz, compared to the observed position of the maximum.

\fig{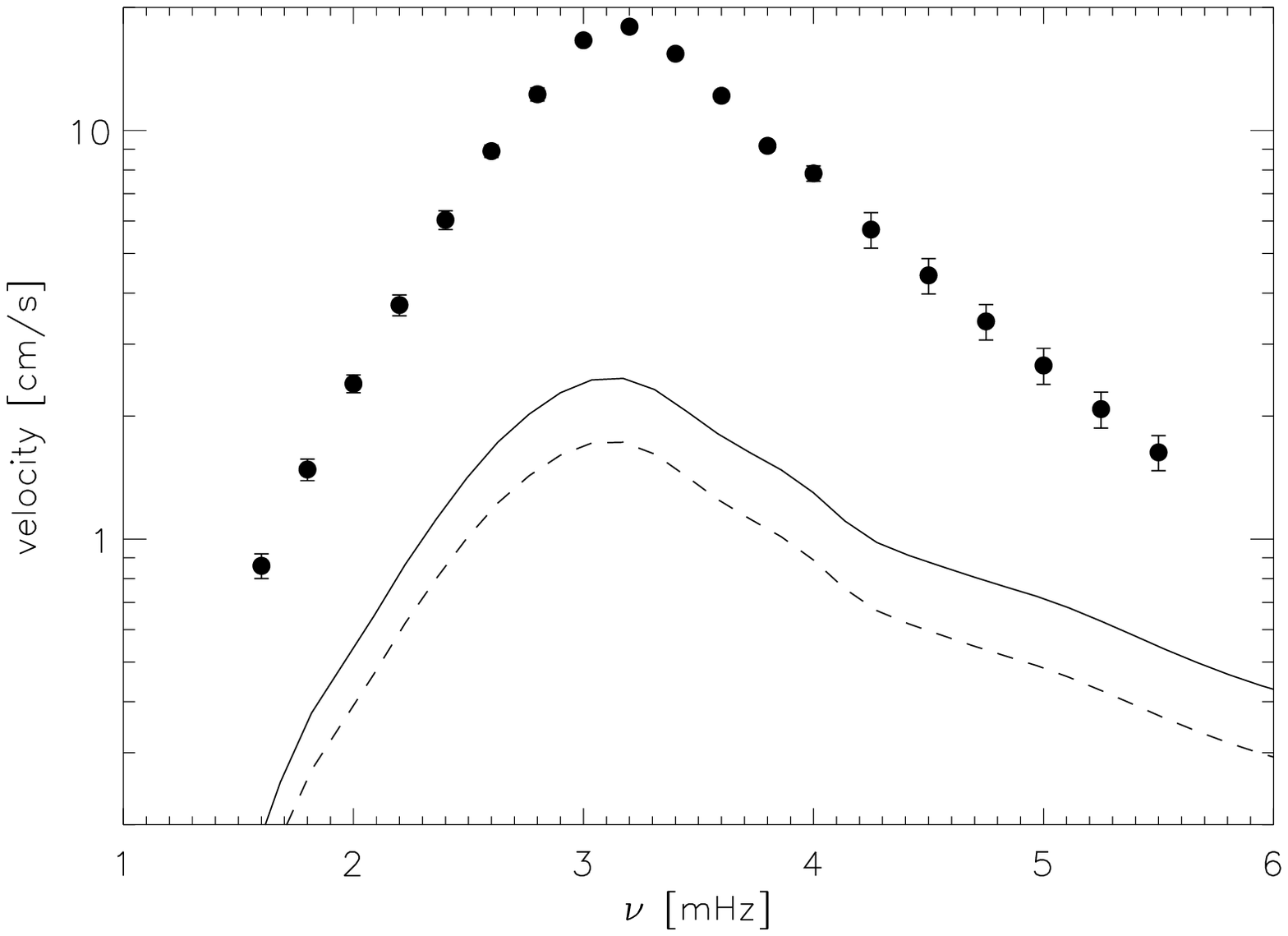}{Root-mean-square velocity $v_s(\omega_0)$: 
dots with accompanying error bars 
represent the solar measurements of \citet{Libbrecht88}.
The continuous curve shows velocity amplitudes computed 
according to Eq.(\ref{eqn:vs}) 
using both the Reynolds and the entropy contributions 
and the dashed curve  using only  the Reynolds stress contribution. A 
Kolmogorov energy  spectrum is assumed. 
Damping rates are deduced from \citet{Libbrecht88} 
and we assumed $\lambda=1$ and $\beta=1.9$. 
 } {fig:Vk}

\section{Impact of different assumptions}

The above calculations were performed 
considering standard assumptions 
for the turbulent spectra and 
the free parameter $\lambda$. We now investigate  
 effects of changing  the turbulent properties, 
 the free parameters $\lambda$ and $\beta$, 
the traitement of the atmosphere and of 
the depth in the atmosphere 
at which the expected surface velocity $v_s$ is computed.

\subsection{Turbulent kinetic energy spectrum} 
\fig{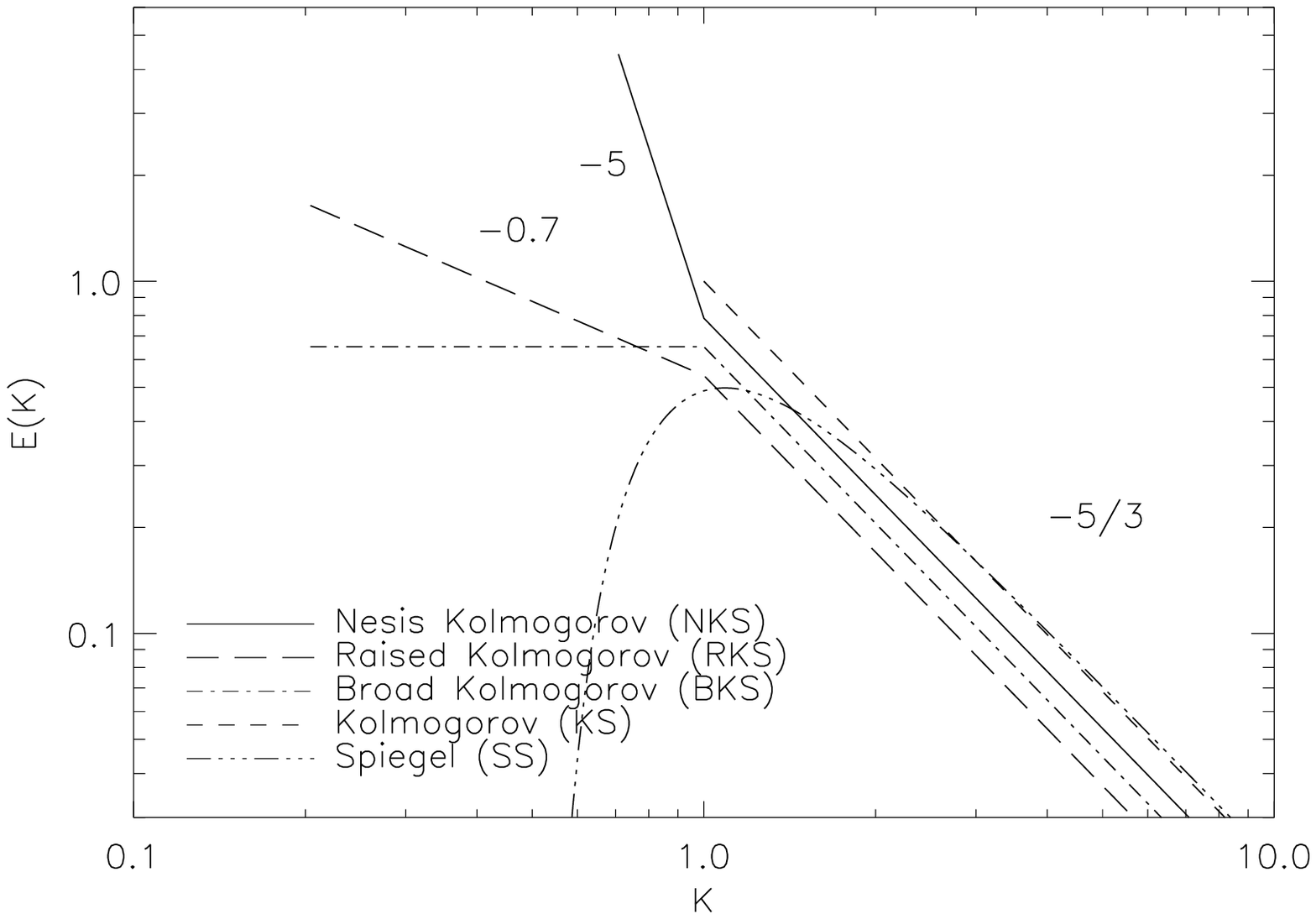}{The RKS, KS, 
SS, BKS and NKS  versus the normalized wavenumber $K$.  
The NKS has  been shifted up  in order 
to distinguish it from the BKS for $K>1$}{fig:spc_cinetique}

In Sect.~2 we  considered the Kolmogorov Spectrum (KS hereafter)
\eqn{
E(K)=  \begin{array}{lcl}
\displaystyle { \frac{u_0^2} {k_0} }  \,  K^{-5/3}  & \textrm{for} & K>1
\label{eqn:kolmo}
\end{array}
}
where $K \equiv k/k_0$, $k_0 = 2 \pi / (\beta  \Lambda ) $ and $u_0^2$ is proportional to the mean square turbulent velocity (see Paper~I). 

Observations of solar granulation show a change of slope  of
the kinetic energy spectrum between the inertial range 
where the spectrum obeys the Kolmogorov law
and the large scale range (small  wavenumbers $k$) corresponding to the energy
injection region.

In order to take large scale eddies into account, we consider first the 
Spiegel spectrum
 (SS hereafter) for comparison with B92 and  \citet{Houdek96}.
\eqna{
 E(K) = \begin{array}{lcl} 
18 \, \displaystyle { \frac {u_0^2} {k_0} } \,  \left ( 2 K \right )^{-5/3} \, J(2 K) & \textrm{for} & K>0.5
\end{array}  
}
with 
\eqn{
J(K) = \left [1-K^{-8/3} -\frac {4} {7}(1-K^{-14/3}) \right ]^2  \nonumber \; .
}
This spectrum takes into account the non-linear interaction 
between turbulent modes of low wavenumber \citep{Spiegel62,Stein67} and is
built for  very large Rayleigh numbers. 

In order to  match  the high-resolution observations of the
solar granulation by  \citet{Roudier86},
 \citet{Musielak94} has proposed the ``Raised Kolmogorov Spectrum''  (RKS hereafter)
\eqna{
E(K) = 
\left \{ \begin{array} {lcl} 
\vspace{0.3cm} 0.54 \, \ds{ \frac{u_0^2} {k_0} } K^{-5/3} & \textrm{for}  & K>1 \\
0.54 \, \ds{ \frac{u_0^2} {k_0} } K^{-0.7}  &\textrm{for}  & 0.2 < K < 1
\end{array}
\right .
\label{eqn:raised_kolmo}
}
and as a possible alternative the ``Broad Kolmogorov Spectrum''   (BKS hereafter)
\eqna{
E(K) = 
\left \{ \begin{array} {lcl} 
\vspace{0.3cm} 0.652 \, \ds{ \frac{u_0^2} {k_0} } K^{-5/3} & \textrm{for}  & K>1 \\
0.652 \, \ds{ \frac{u_0^2} {k_0} }  &\textrm{for}  & 0.2 < K < 1 \; .
\end{array}
\right .
\label{eqn:broad_kolmo}
}
We also consider  the ``Nesis Kolmogorov Spectrum'' (NKS, hereafter)
\eqna{
E(K) = 
\left \{ \begin{array} {lcl} 
\vspace{0.3cm} 0.655 \, \ds{ \frac{u_0^2} {k_0} } K^{-5/3} & \textrm{for}  & K>1 \\
0.655 \, \ds{ \frac{u_0^2} {k_0} } K^{-5} &\textrm{for}  & 0.7 < K < 1
\end{array}
\right .
\label{eqn:nesis_kolmo}
}
which mimics the solar kinetic spectrum observed by \citet{Nesis93}.

The above five kinetic energy spectra are plotted in 
Fig. \ref{fig:spc_cinetique}. All these 
spectra exhibit the same  $-5/3$ slope 
at large wavenumbers. 
The RKS, the BKS, the SS and the NKS 
take into account large-scale eddies 
lying in the injection region.

 As stressed by \citet{Rieutord00}, differences at low wavenumber between the kinetic spectra obtained from observations of the solar granulation are a consequence of uncontrolled data-averaging procedures. Properties of the kinetic spectrum at low wavenumber (mesogranulation) are thus not yet well depicted by the observations of the solar surface. In the present paper we nevertheless study the different spectra of Eqs.(\ref{eqn:raised_kolmo}-\ref{eqn:nesis_kolmo}) in order to gauge the sensitivity of the acoustic power emission to the properties of the turbulent spectrum in the injection region.

Fig. \ref{fig:VRS_spc}  presents  amplitudes computed with 
the sole Reynolds stress contribution and for the different kinetic energy 
spectra of Fig. \ref{fig:spc_cinetique}. 
Only small differences in  the velocity amplitudes are observed arising from
different spectra at low frequencies.  The main differences are observed at high
frequency
where the RKS and  BKS exhibit a steeper slope 
than the KS and SS. These features are explained below.

\fig{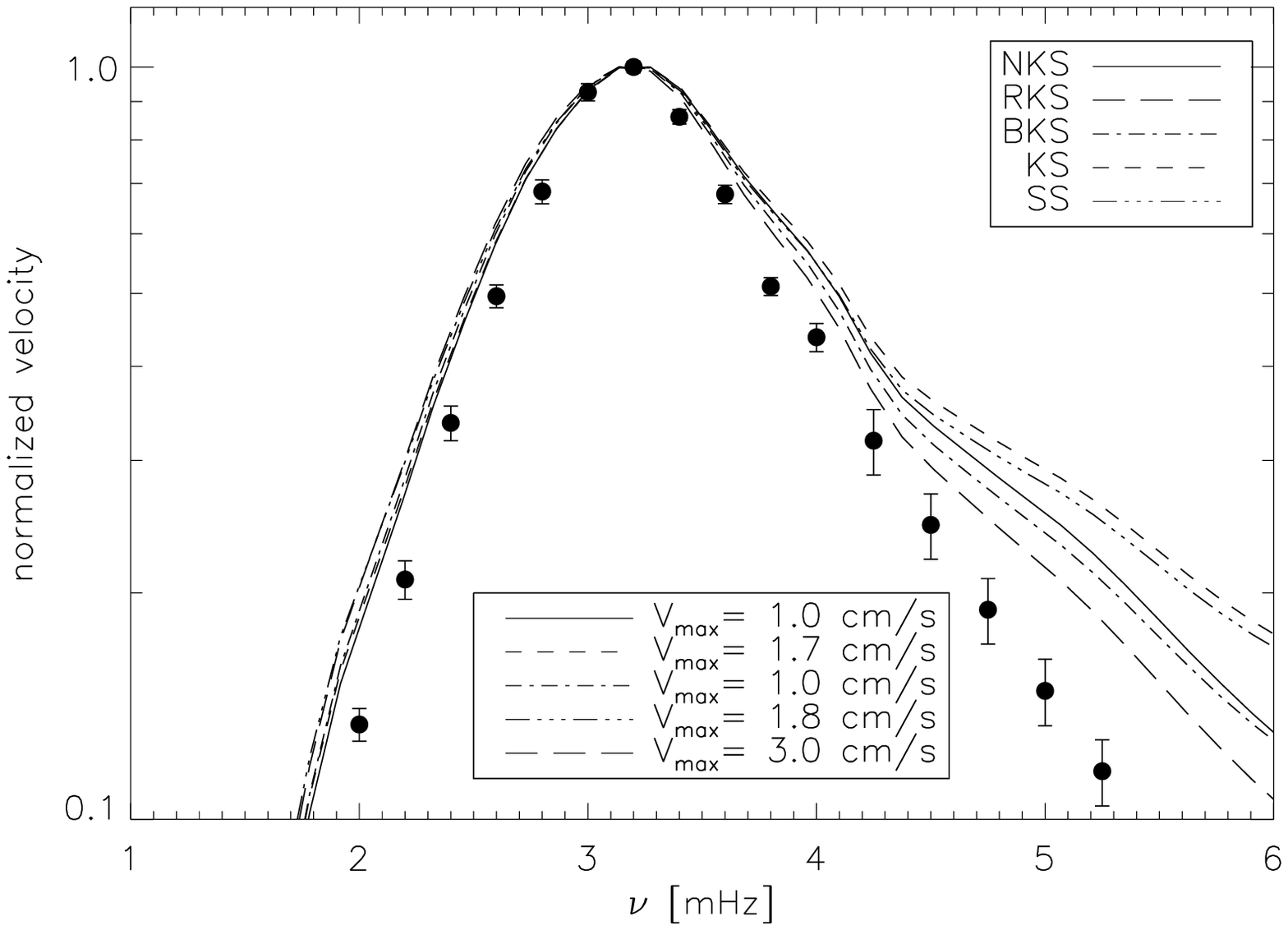} {Amplitudes computed
 with  the Reynolds stress contribution
only  and for the kinetic energy spectra of Fig. \ref{fig:spc_cinetique}.  
A gaussian time spectrum has been adopted. 
The curves are normalized to their maximum value 
and have intentionally  been shifted by $+0.1$~mHz in order 
to match the position of the observed maximum. 
All other assumptions are those of Fig. \ref{fig:Vk}. 
} {fig:VRS_spc}

The velocity amplitudes obtained by B92  using  the KS  are  shifted  by a
larger amount toward  lower frequencies compared to ours and 
differences between the KS and SS are found to be 
larger in   \citet{Houdek96}, in which 
computations are  based on the formulation of B92.
The last feature is due to the way B92  implemented the SS and 
the first feature is due to the way we  normalized 
the turbulent energy spectra (Eq. 63 in Paper~I).

With our formulation, velocity amplitudes using the SS are similar to
those using  the  KS. Therefore we no longer consider  the SS  in the
remaining part of this work.

From here on, amplitudes calculations include both Reynolds stress  and
entropy contributions.
Fig. \ref{fig:VRSEP_spc} 
shows amplitudes computed with the 
free parameters  set to $\lambda=1$, $\beta=1.9 $ with
$\Phi=2$.
 Large differences are found at 
high frequencies arising from the use of different  turbulent spectra. 
The RKS leads to larger amplitudes and fits better 
the solar observations at high frequencies.
At low frequencies, the RKS overestimates $v_s$ more 
than  the estimates including only the Reynolds stress contribution. 
The RKS contains a significant amount of 
kinetic energy at low wavenumbers. 
At low frequencies the excitation process is 
mainly due to the large scale eddies. Furthermore it is
stressed in Paper~I that the entropy contribution is 
more efficient for long-period oscillation modes. 
Therefore the RKS injects through the entropy contribution 
a larger amount of energy to the low frequency modes than the other spectra. 
For the other spectra we found few differences in terms 
of frequency dependence between computations including 
  only the Reynolds stress contribution and including 
both Reynolds stress  and entropy contributions.

\fig{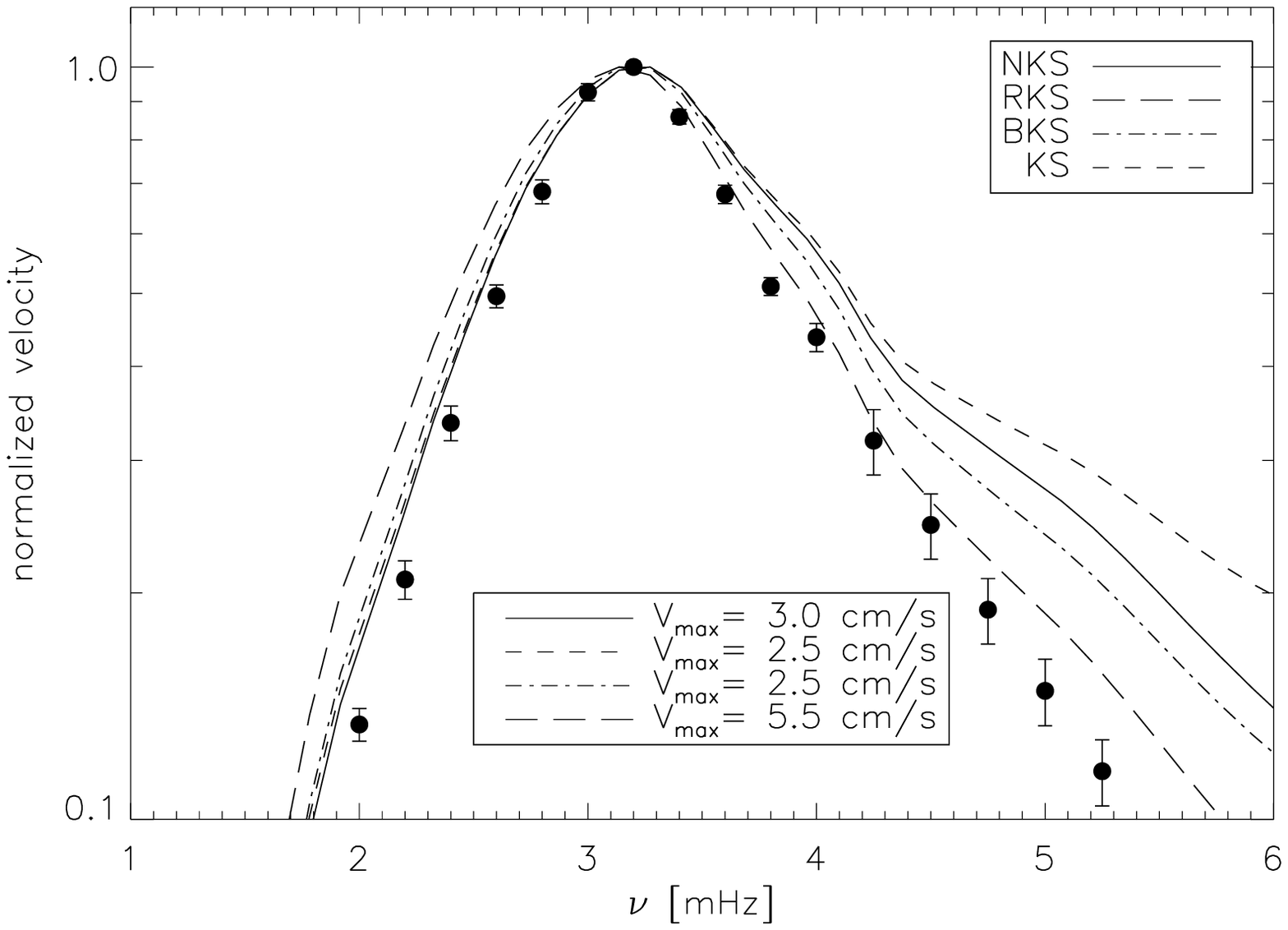} {Velocity amplitudes computed 
with  both the Reynolds and the entropy contributions and 
for different kinetic energy spectra. A gaussian time spectrum has been
adopted. The curves are normalized to their maximum value
and have intentionally  been shifted by $+0.1$~mHz in order
to match the position of the observed maximum. Dots are observational points
as defined in Fig.\ref{fig:Vk}. The free parameters are set to $\lambda=1$, $\beta=1.9$.}
{fig:VRSEP_spc}

The computed $P_{\omega_0}$ is plotted in Fig. \ref{fig:pRSEP_spc} in order
to emphasize the changes due to the use of different turbulent spectra.   
Also shown is the power $P_{\omega_0}$ derived from 
the observations according to Eq.(\ref{eqn:vs}) 
where the radius $r_s$ corresponds to the layer at which $v_s$ is measured.
We observe large differences between the oscillation  spectra  $P_{\omega_0}$. 
At high frequency, the KS strongly overestimates  $P_{\omega_0}$ 
while the RKS and the BKS are closer to the oscillation power  
derived from the observations. Thus the RKS and BKS inject 
relatively less energy into the high frequency oscillations. 
This is explained as follows:   contributive eddies are 
those for which  $ \tau_k \omega_0 \lesssim 1$.  
At a given wavenumber $k$ the RKS induces a smaller 
velocity $u_k$ for the considered eddy. 
As $\tau_k \approx (k u_k)^{-1}$ for a given oscillation frequency, 
the RKS then involves eddies with larger wavenumbers than the KS.  
Eddies with a high wavenumber carry less energy and consequently 
  the RKS injects relatively less  power into the oscillations 
of high frequencies than the KS does.

At low frequencies smaller differences are observed between the spectra 
because for small values of the oscillations 
frequencies all the eddies  of the spectra
 are involved in the excitation process. 
However, the RKS overestimates $P_{\omega_0}$ while 
the BKS leads to an intermediate result between the RKS and the KS.

\fig{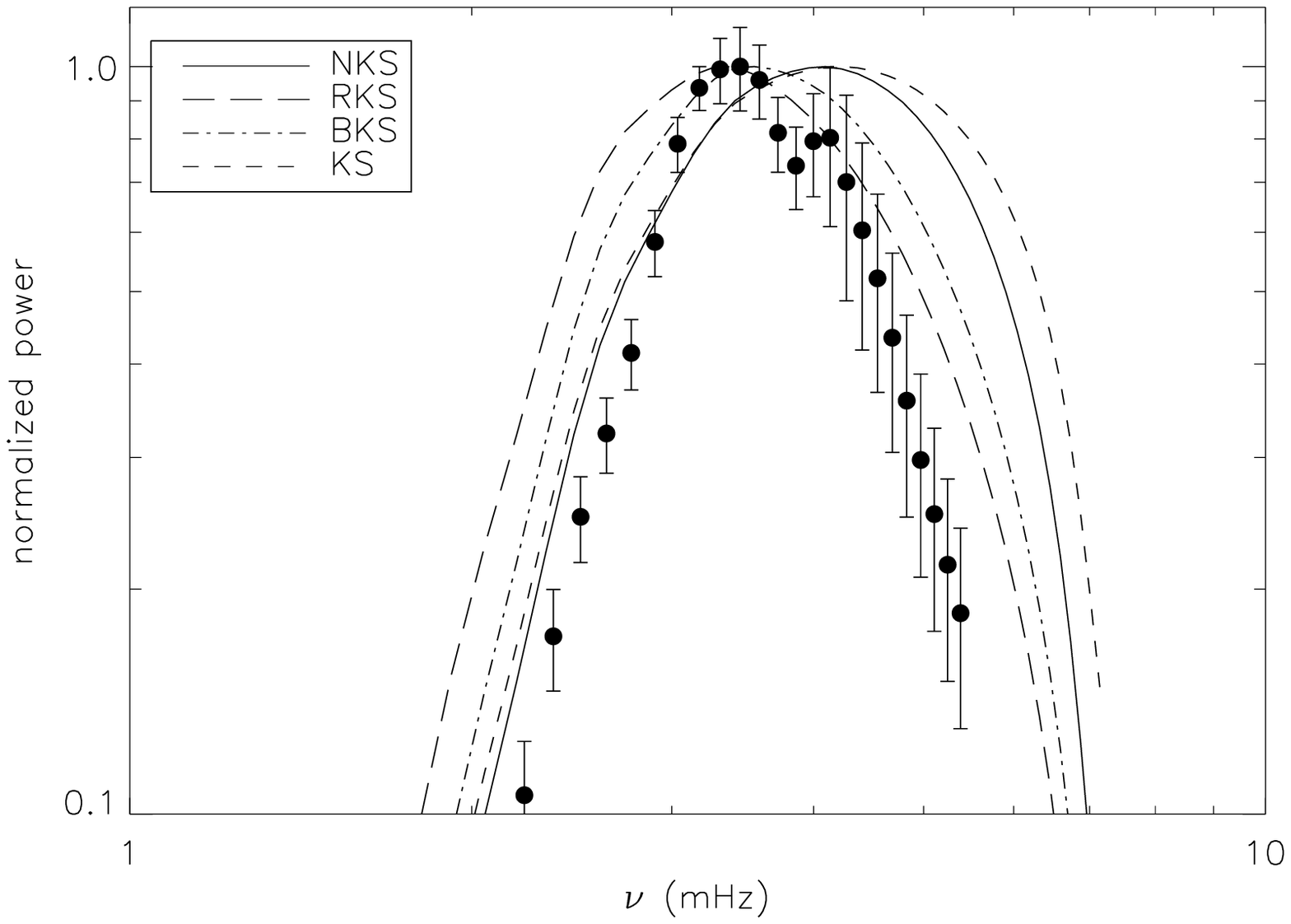} {$P_{\omega_0}$ 
 for different kinetic energy spectra are presented in $log-log$
 representation.  
Dots represent the oscillation power derived from the solar observations 
with the help of Eq. (\ref{eqn:vs}). Corresponding errors bars are plotted. } {fig:pRSEP_spc}

\begin{figure*}[tbp]
	\resizebox{\hsize}{!}{\includegraphics{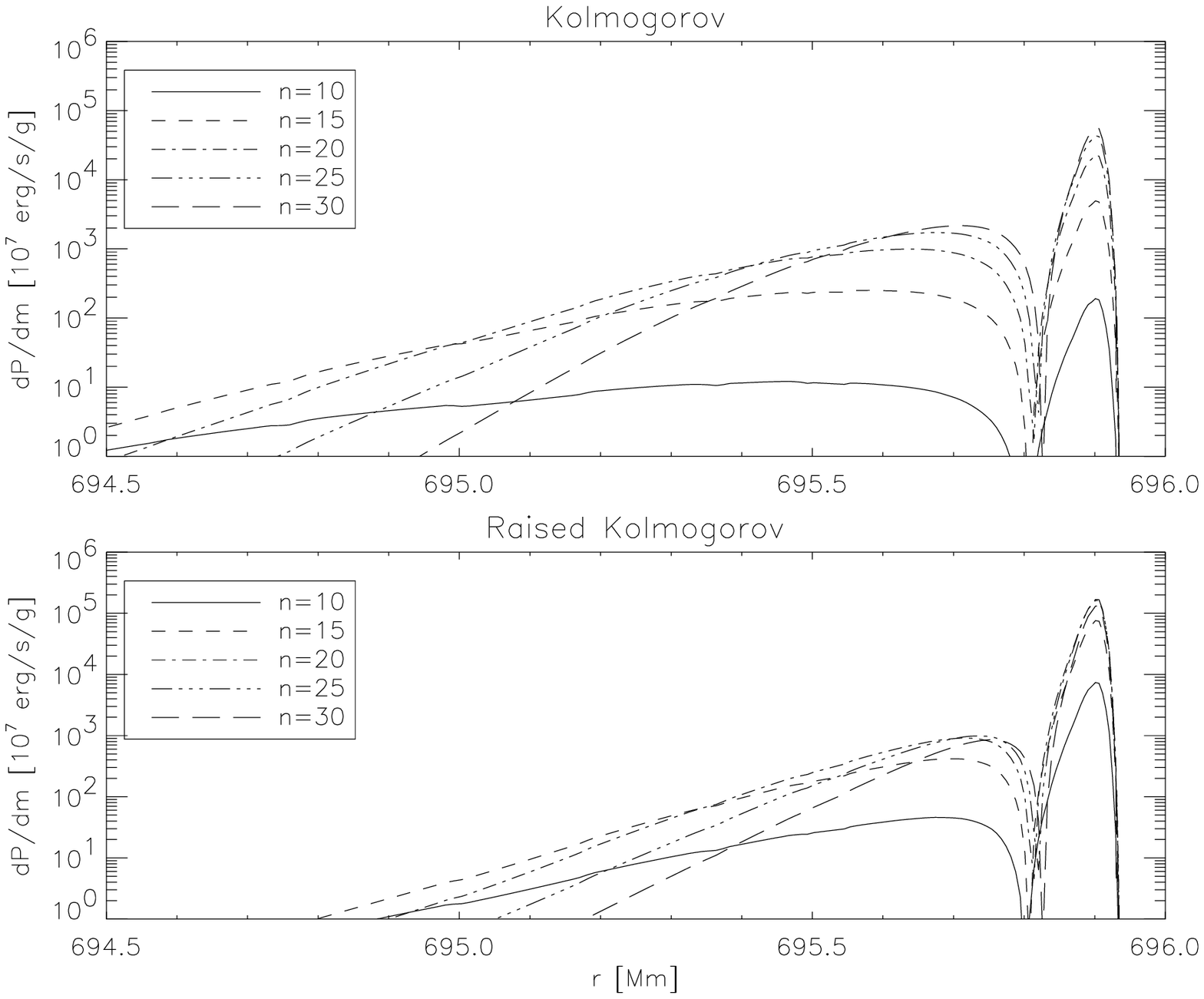}}
\caption{Depth dependence of the injected power  in the solar outer layers
computed for the oscillation modes of radial 
order $n=[10,15,20,25,30]$. The abscissae are the stellar radius from center  in Mm. 
The oscillation power is plotted for the KS (top) and  
the RKS (bottom) with $\lambda=1$ and $\beta=1.9$. All other assumptions are the
same as in Fig.4.
 }
\label{fig:dP_dmRSEP}
        \end{figure*}

Fig. \ref{fig:dP_dmRSEP} compares  the sizes of the excitation region
 when  using the KS and the RKS for several oscillation modes. 
It shows that the depth dependence of $P(\omega)$ behaves according to the depth
 dependence of the associated displacement eigenfunction. For higher radial
 order, the mode is more concentrated toward the surface and consequently 
 the excitation region is shallower.
The extension of the excitation region 
is about twice the size with the KS than with the RKS; i.e. $\approx 1.5$~Mm 
with the KS against $\approx 0.8$~Mm with the RKS and for a mode of order
 $n=20$ 
and a  frequency $\nu \approx 3$~mHz. 
This can be explained as follows: For a given wavenumber $k$, 
the correlation time of an eddy, is larger with the
 RKS than with the KS because the RKS induces smaller velocity 
$u_k$ at a given wavenumber $k$. 
Therefore the region where  $ \tau_k \omega_0 \lesssim 1$ is
  smaller with the RKS. This explains the strong differences 
in terms of depth of excitation found between the
 KS and the RKS (Fig. \ref{fig:dP_dmRSEP}). The BKS induces 
an intermediate depth of excitation between the RKS and the KS (not shown here).
Therefore the RKS induces a thinner excitation region which 
is closer to the one  found in the simulation
 performed by  \citet{Stein00II}.


\subsection{Turbulent entropy spectrum}

It has been assumed above that the entropy spectrum  
exhibits the same behavior as the kinetic energy spectrum. 
However,  in paper I, we stress that the entropy 
spectrum exhibits the same behavior as the temperature  spectrum and 
 the turbulence theory \cite[Chap VI-10]{Lesieur97} 
predicts a specific  spectrum for the turbulent temperature fluctuations which 
derives from the scalar nature of the temperature.   
 We therefore proposed the ``conductive'' entropy spectrum  
derived from  the theoretical spectrum expected for the temperature fluctuations \citep{Lesieur97}
\eqna{ 
E_s (K) = \left \{ 
\begin{array}{lll} 
\displaystyle { \frac{a_0 \tilde{s}^2} {u_0^2 k_0} E(K)  } & \textrm{for} & 1<K<K_C \\
\displaystyle {  \frac{a_0 \tilde{s}^2} {u_0^2 k_0} } \left ( \frac{K_C} {K} \right )^4 E(K)   & \textrm{for} & K_C<K 
\end{array}
\right .
\label{eqn:E_s}
}
where $a_0$ is a normalization factor, $E(K)$ is the kinetic energy spectrum, $\tilde{s}^2$ is the entropy scalar variance (Eq. 88 of Paper~I) and  $K_C \equiv k_c / k_0 $ where  $k_c$ is the wavenumber which separates the inertial-convective range  from the inertial-conductive range.
When the kinetic energy spectrum $E(k)$ obeys the Kolmogorov scaling law, the  entropy spectrum verifies
\eqna{ 
E_s (K) = \left \{ 
\begin{array}{lll} 
\displaystyle {   \frac{a_0 \tilde{s}^2} {k_0} K^{-5/3}  }& \textrm{for} & 1<K<K_C \\
\displaystyle {   \frac{a_0 \tilde{s}^2} {k_0} K_C^4 \,  K^{-17/3} } & \textrm{for} & K_C<K  \; .
\end{array}
\right .
\label{eqn:E_s_kolmo}
}
This spectrum, that we  call the ``Kolmogorov Conductive spectrum'' (KCS,
hereafter), 
matches observations of intensity fluctuations 
of  the solar granulation  
performed by   \citet{Espagnet93}  and \citet{Nesis93}.

There is no theoretical constraints on the  value of $K_C$ ;
therefore $K_C$ may be considered as a free parameter of our formulation. 
We thus  investigate different values of $K_C$ with the use of the KCS (Eq.15). 
The results are presented in Fig. \ref{fig:VRSEPkc}.

\fig{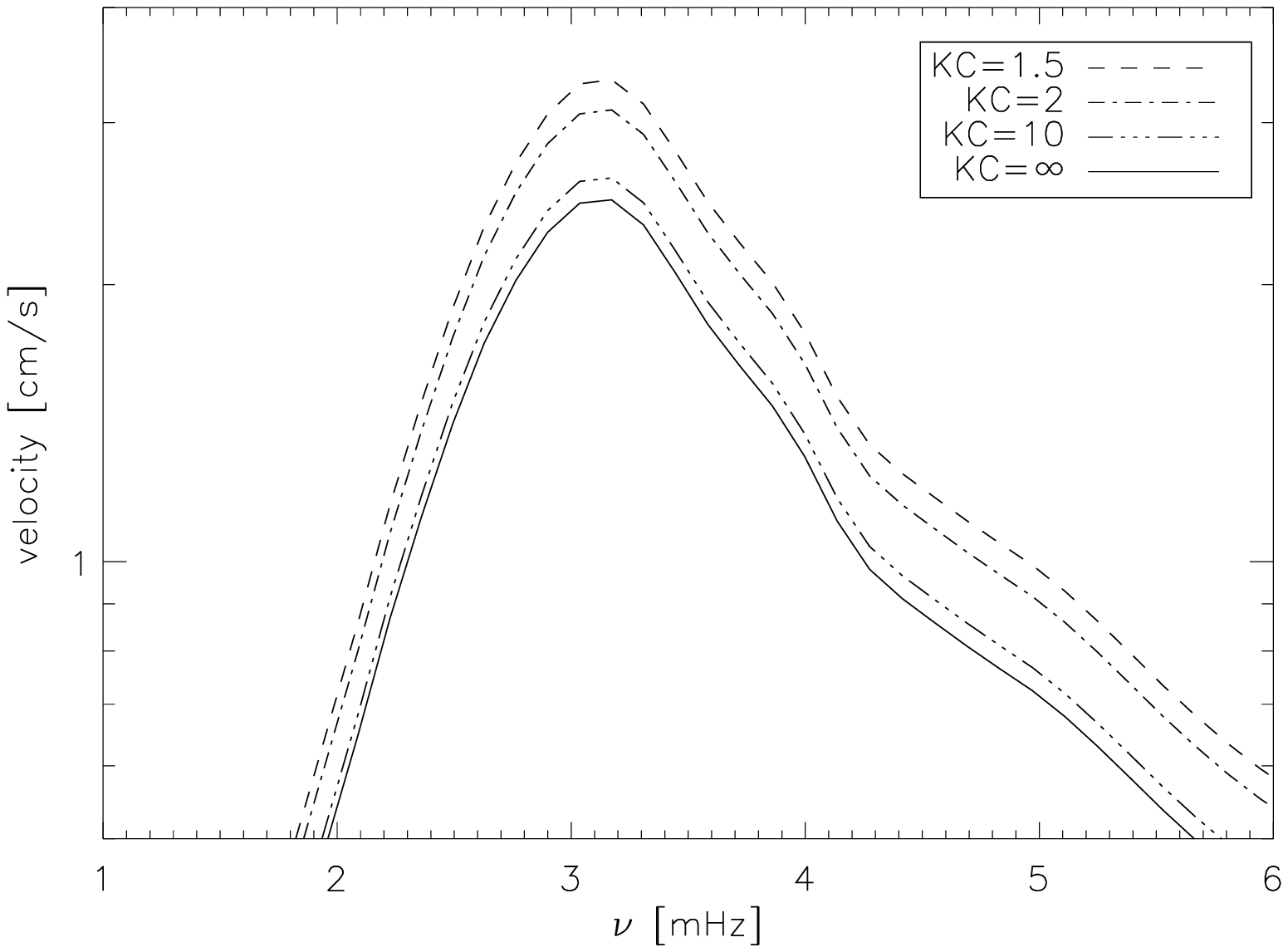} {Amplitudes computed with the ``Kolmogorov Conductive  
spectrum'' (KCS)  for different values of $K_C$.  
All other assumptions are those of Fig.4.  
 } {fig:VRSEPkc}

We observe no change in the shape when varying $K_C$, only a decrease of the global 
amplitudes  with increasing  $K_C$. 
As $K_C$ decreases, the major part of the energy in 
the entropy energy spectrum is concentrated at low wavenumbers. 
Eddies with small wavenumbers carry more energy 
 than those with larger wavenumbers. Therefore a
decreasing value of $K_C$ induces an increase of 
$P_{\omega_0}$  and thus the shape of $v_s(\omega_0)$ 
remains unchanged. We found similar results 
with the entropy ``conductive'' spectra based on the  RKS, BKS and NKS
 (not shown here) and derived from Eq.(\ref{eqn:E_s}). 

Therefore we cannot constrain the value of $K_C$ from the seismic solar observations. 
However  \citet{Espagnet93} shows that 
the inertial-convective domain of the temperature 
spectrum starts from granular scales of
 about $k_0 \approx 3 \, \mathrm{M^{-1}m}$ while  
the inertial-conductive range starts from  $k_c \approx 6 \, \mathrm{M^{-1}m}$.
 This leads to a value $K_C=2$.

In the remaining part of this work,  the computation of $P_{\omega_0}$ 
 will be performed assuming the ``conductive'' spectra for the entropy
 contribution with $K_c=2$.

\subsection{Time spectrum}

The time spectrum function of eddies with given $k$ is 
usually modelled by a  gaussian function in which linewidth 
is proportional to the time correlation of the eddy 
\cite[Paper~I]{GK77,Balmforth92c}. As proposed by   \citet{Stein67} 
and  \citet{Musielak94} 
we also consider below other forms  
for the time spectrum $\chi_k (\omega )$ such  as the modified gaussian form
\eqn{
\chi_k (\omega ) = 
 \frac{2}{\sqrt{\pi}} \frac{\omega^2 }{\omega_k^3} e ^{-(\omega/\omega_k)^2 } 
\label{eqn:modified_gaussian}
}
and an exponential form
\eqn{
\chi_k (\omega ) = 
\frac{1}{2 \omega_k}  e ^{-| \omega/\omega_k | } \; .
}
The quantity $\omega_k$ controls the width of the eddy time spectrum and is therefore related to the eddy  correlation time as $\omega_k= 2 / \tau_k$ (see Paper~I).
All time spectra are normalized such that 
\eqna{
\int_{-\infty}^{+\infty}  \chi_k (\omega) d\omega = 1 \; .
}
This choice of normalization differs from the one
 considered by   \citet{Musielak94}, but 
 is in aggrement with  \citet{Tennekes82} requirements. 

Fig. \ref{fig:VRSEPrk_chi} shows 
the normalized velocity amplitudes $v_s$ computed with 
 the RKS and for different time spectra :  the gaussian form, the modified gaussian 
and the exponential forms.
Both the exponential and the modified gaussian forms yield amplitudes which are
too large compared with the
observations at high frequencies while  
the gaussian form gives rise to amplitudes which are  closer to the
observed amplitudes.
At low frequencies both the modified gaussian and the exponential forms 
fall slightly closer to the observations than the gaussian form.
In contrast, computed $v_s$ assuming the KS leads to the opposite behavior 
for the gaussian form, {\it i.e.} at low frequencies, 
computed $v_s$ with the gaussian form are closer  
to the observations than the other forms (not shown here).

The strongest disagreement between observed and computed surface velocity
amplitudes using KS and  RKS is obtained at high frequency for the exponential form.
Therefore we discard this time spectrum from now on.
We cannot  clearly discriminate between the 
modified gaussian form and the gaussian form and  choose to adopt the gaussian form.

\fig{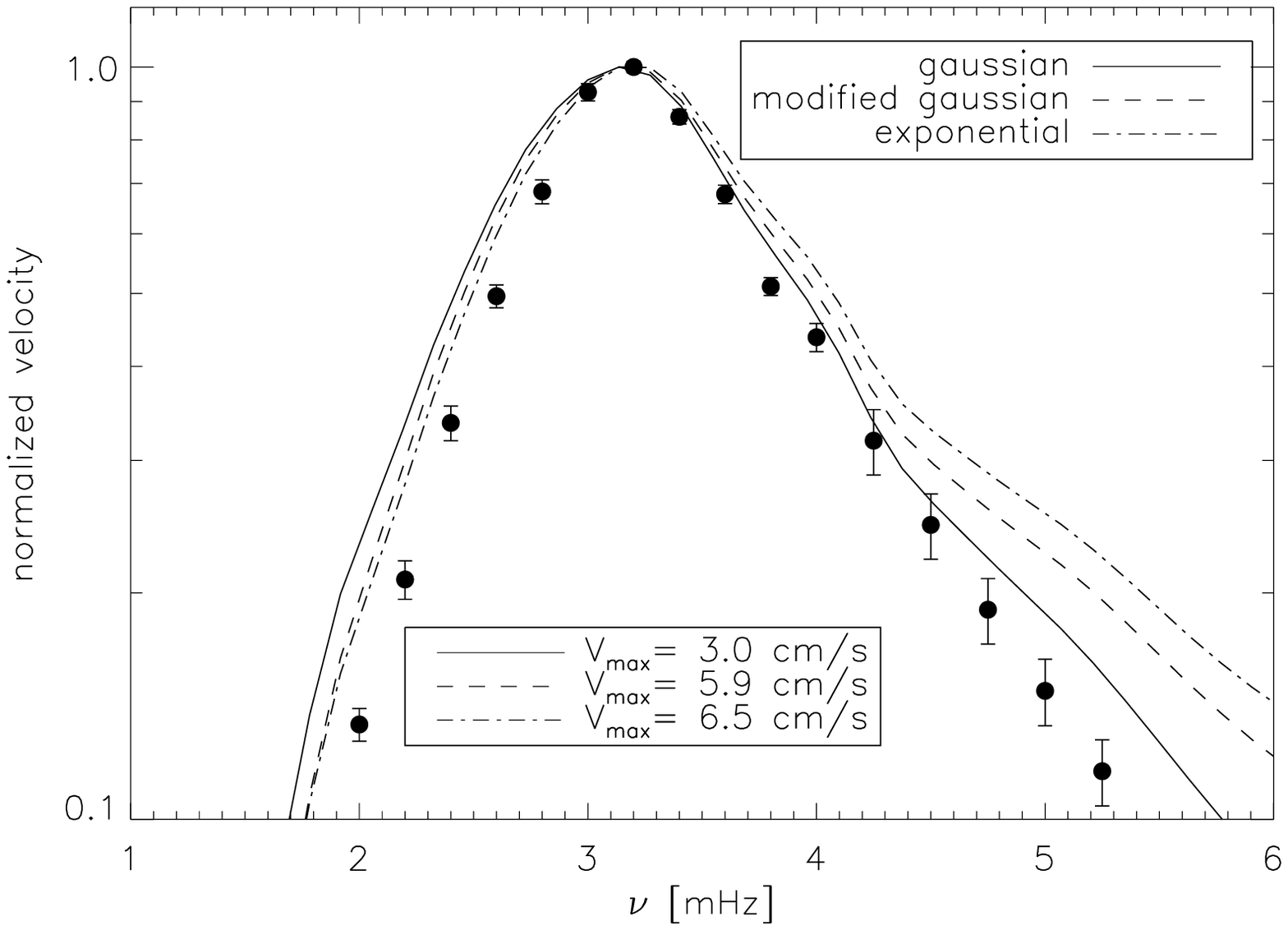} {Computed $v_s(\omega_0)$ assuming the RKS  is
plotted for 
different frequency 
factors $\chi_k (\omega )$. 
All other assumptions are those of Fig. \ref{fig:VRS_spc} }{fig:VRSEPrk_chi}

\subsection{Influence of the atmosphere}

We have computed two  solar models, one with the Eddington classical gray atmosphere 
and one with the Kurucz ATLAS 9 \citep{Kurucz91} solar model atmosphere. 
The differences in the computed powers (not shown here)  
are found to be  much smaller than effects due to the choice 
of the kinetic spectrum and than the errors in the observations.

On the other hand, the measured velocity amplitudes $v_s$ depend on 
the height in the atmosphere 
where the observations are sampled. Fig. \ref{fig:VRSEPrk_hatm} 
shows  theoretical velocity amplitudes calculated at different heights in the atmosphere.
\fig{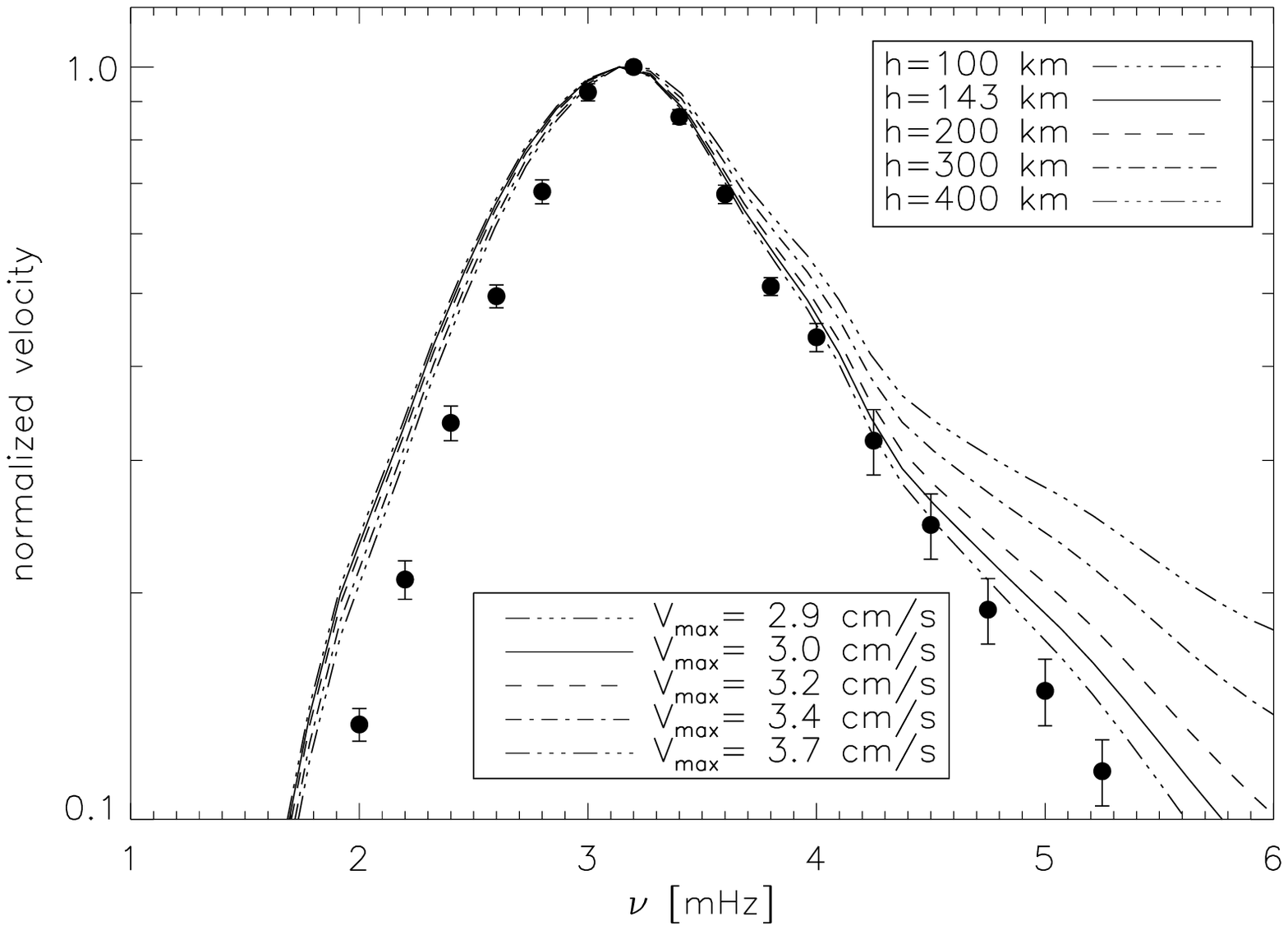} {Velocity amplitudes $v_s$ computed at different
levels in the atmosphere and using  the RKS. 
Other assumptions are those of Fig.4.
}{fig:VRSEPrk_hatm} 

Changes in the eigenfunction profiles at the solar surface 
are larger with high frequency modes. 
Therefore increasing the  height in the atmosphere 
where $v_s$ is computed  leads  to larger changes at high frequencies 
($\nu \gtrsim 3.5$~mHz).
Almost no effect  is observed at low frequencies. 
This shows  that the physics of the excitation process
 is more easily constrained  by the  observations at low frequencies ($\nu \lesssim 3.5$~mHz).

\subsection{Effects of the free parameters $\lambda$ and $\beta$}
\label{sec:effects_parameter}

With $k_0$ given by Eq.6, the  eddy time correlation 
length (Eq.5) can be related to $\beta$ and $\lambda$ as
\eqn{
\tau_k= \frac{ \lambda \, \beta }{\omega_\Lambda K u_K }
}
where $\omega_\Lambda= 2 \pi u_0 / \Lambda$ and $u_K= u_k/u_0$. 
Thus both $\beta$ and $\lambda$ are involved in the 
expression of the amplitude (Eq. 101, Eq. 102 and Eq. 103 of Paper~I). 
The product $\beta \, \lambda$ controls the frequency-dependence of the
oscillation power. Fixing the value of $\beta$ to $1$, 
we investigate the effects of changing  the parameter  $\lambda$. 
In terms of frequency dependence, this is equivalent 
to fixing $\lambda$ and to varying the  parameter $\beta$.

Fig. \ref{fig:VRSEPk_lambda}  shows  velocity amplitudes $v_s$ computed 
when assuming the KS, the RKS and the NKS for different  values of  $\lambda$.
The value of $\lambda$ (or $\beta$) controls the extension of the excitation 
region through $\tau_k$.
 Small values of $\lambda$ lead to a larger excitation region 
and therefore favor long-period oscillation modes. This explains the larger 
 frequency  shift of the maximum amplitude between the computations 
and the observations obtained with   $\lambda \lesssim 1$. 
Changing $\lambda$ has smaller effects at low frequencies than at high frequencies 
because at low frequencies all the eddies in the spectra
 are involved in the excitation process and 
the frequency dependence is mainly controlled by the 
mode inertia and the derivatives of the eigenfunctions. 
Changes are nevertheless larger when using  the RKS than the NKS and the KS.

For $\lambda=0.5$ and using the KS,
 the frequency shift between the numerical results  and the observations 
is about $0.25$~mHz and the computations overestimate the amplitudes at low
 frequencies. In that case,  the velocity $v_s(\omega_0)$   is close that 
 found by B92 and  \citet{Houdek96} 
which assumed $\lambda=1$. 
This result is consistent with the  \citet{Osaki93} statement that the B92
 formulation leads to a  source of excitation that is too extended.
Our estimates with $\lambda=1$ and using the KS are closer 
to the observations than those performed by B92 and  \citet{Houdek96}. 
This difference is due to the way we have normalized the  turbulent energy spectra.
Therefore our formulation leads to a  smaller depth of
 excitation  and is in better agreement with the observations.

We observe large changes at 
high frequencies with $\lambda$ 
for all the spectra. 
Increasing $\lambda$ induces a steeper slope at high frequency. 
This can be explained as follows: contributive eddies are 
those for which  $ \tau_k \omega_0 \lesssim 1$. 
These eddies are the eddies for
 which $k \gtrsim k_0 ( \lambda \beta \, \omega_0/ K u_K
\omega_\Lambda)^{3/2}$. 
As $\lambda$ increases, only eddies with higher wavenumbers $k$ are 
involved in the excitation process. Therefore, because 
eddies with high wavenumbers carry less energy, 
$P_{\omega_0}$   is smaller with increasing value of $\lambda$.

\fig{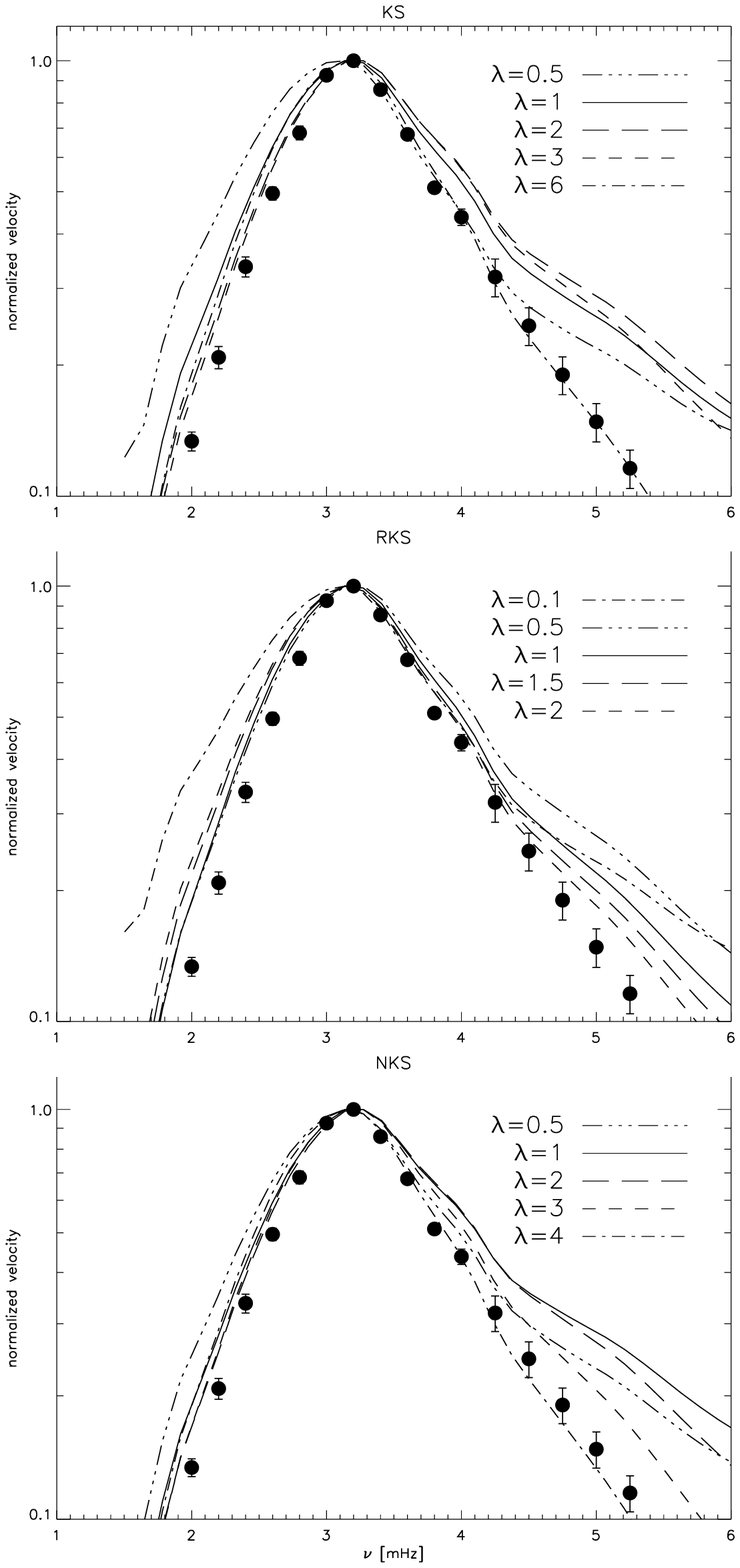} {Surface velocity $v_s(\omega_0)$ computed 
with $\beta=1$ for different values of the parameter $\lambda$. 
The top panel concerns the KS, the middle one the RKS and the bottom one the NKS.
All other assumptions are those of Fig. \ref{fig:VRSEP_spc}. }{fig:VRSEPk_lambda} 

\vskip 0.3truecm
The oscillation amplitude increases  as $\beta^4$, hence 
 $\beta$ plays a major role in the control of the 
maximum amplitude. A theoretical constraint can be obtained  for $\beta$
as $\beta \lambda \lesssim 2.7 \,  \Phi^{1/2}  $ (see paper I). 
For $\Phi=2$, we have $ \beta \lambda \lesssim 4$.

\section{Solar calibration}
\label{sec:Solar calibration}

We need to determine the  free parameters  $\lambda$ and $\beta$ by 
 calibrating the theoretical amplitudes on the solar observational ones.
For each of the various kinetic and ``conductive'' spectra,
   we first adjust $\beta \lambda$  in order to match as closely as possible the position of the maximum and when possible
the frequency dependence of $v_s(\omega_0)$ to  the
observed surface velocity given by   \citet{Libbrecht88}.
We next adjust $\beta$ and $\lambda$ in such a way as to obtain
the observed maximum value  of $18\,{\rm cm\,s}^{-1}$.
Table \ref{tab:adjusted_parameters} gives the values of the adjusted parameters.  

Fig. \ref{fig:VRSEP_cspc_fit} shows amplitudes computed 
with the RKS, KS, BKS and NKS with $\beta$ and $\lambda$ set to their
adjusted values.  The KS and NKS  yield  much better fits to 
the seismic observations than the other spectra. 

Observations of solar granulation by  \citet{Espagnet93} 
show that $k_0 \simeq 3 \, \mathrm{Mm^{-1}}$ whereas a value of $k_0 \simeq 7\, \mathrm{Mm^{-1}}$  can be inferred from the observations of  \citet{Nesis93}. 
These values correspond to $\beta \approx 8$ and 
$\beta \approx 4$ respectively. 
The values of $\beta$ adjusted with the RKS, the BKS and the NKS  
are closer to the value of $\beta$ suggested by the observations 
of \citet{Nesis93} while the value 
suggested by the observations of  \citet{Espagnet93} favors the KS.
Therefore $\beta$ is not well constrained by the observations of the solar 
granulation and this cannot help to discriminate between  the different
turbulent spectra. 

The best fits are obtained with the NKS and the KS.
The power computed with the KS seems to fit 
best  the observations; this can be explained by the existence of 
compensating effects. We indeed know from the observations of the solar 
granulation 
that kinetic energy is carried by large-scale eddies. This corresponds to
 the $k-$injection region lying at scales above  the inertial range,  which is
not described by KS.  Furthermore, the theoretical constraint imposed on $\beta \lambda$ 
(see Sect.~\ref{sec:effects_parameter}) suggests that 
the adjustment value of $\beta \lambda$ found by fitting 
the amplitudes with the KS to the observations 
is not valid, while the other spectra verify this constraint.
 One is therefore left with the NKS, $\lambda=0.76$ and
$\beta=5.0$ as our best fit of the solar data.
However, the fit is not perfect at low frequencies. We show in  \citet{Samadi00III} that this is related to the treatment of the static model and the corresponding oscillations.

\begin{table}
\caption{Values of the adjusted  parameters $\beta$ and $\lambda$. 
The wavenumbers $k_0$ of the largest eddies in the inertial range 
have been evaluated at the solar radius  
according to the relation $k_0 = 2 \pi / \beta \Lambda$. 
We have assumed  the ``conductive'' spectra and $K_C=2$.}
\label{tab:adjusted_parameters}
\begin{center}
\begin{tabular}{lllll}  
spectrum  &   $ \beta \lambda$   & $ \beta $ & $ \lambda$ & $k_0 \, (\mathrm{Mm}^{-1}) $ \cr \hline
RKS & 1  & 3.53 & 0.28  & 7.3 \cr
BKS & 1  & 4.44 & 0.22 & 5.8 \cr
NKS & 3.8 & 5.00 & 0.76 & 5.1 \cr
KS & 5.7 & 9.51 & 0.60  & 2.7 \cr
\end{tabular}
\end{center}
\end{table}

\fig{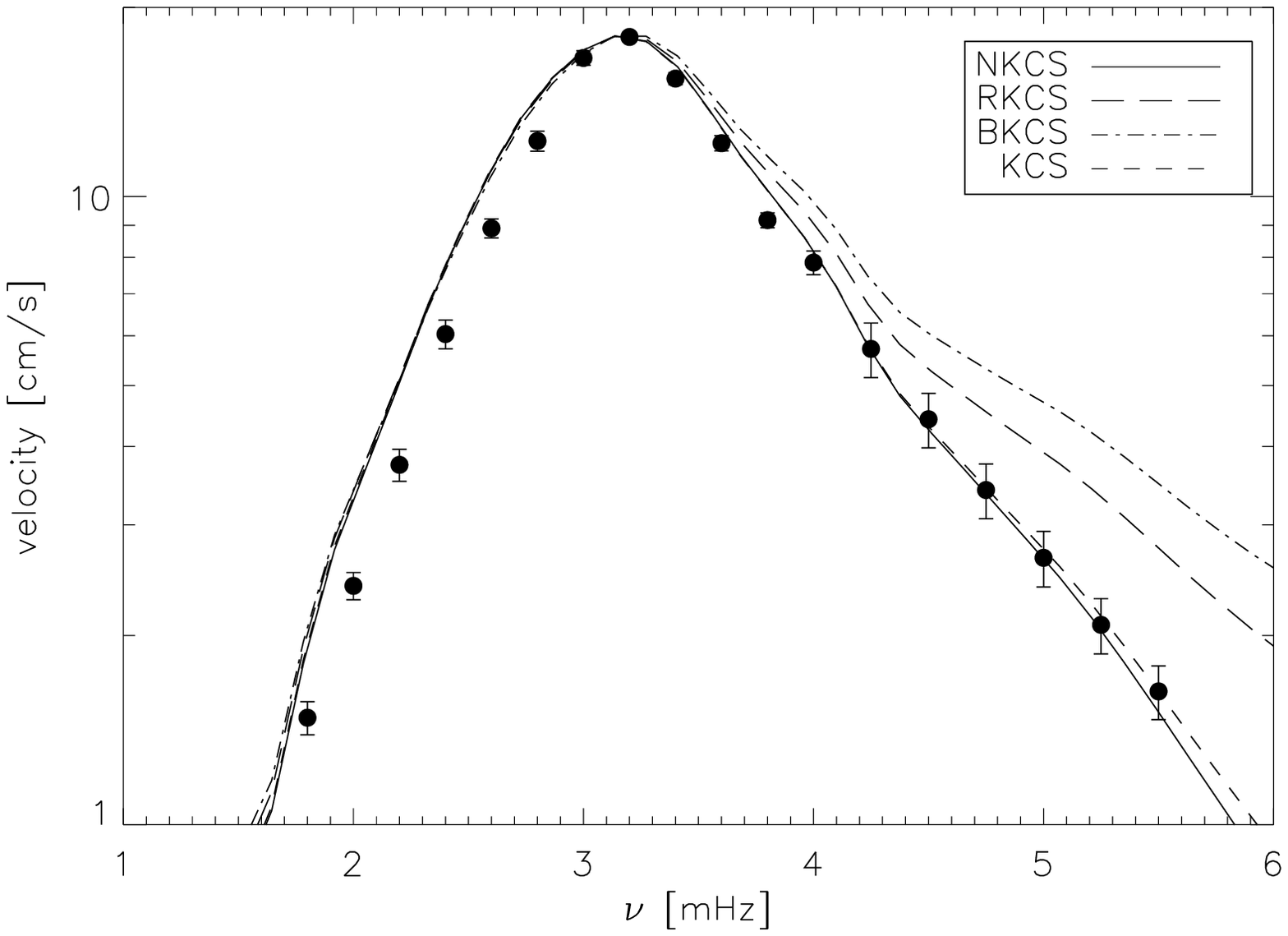} {Velocity mplitudes computed  with the RKS, BKS, NKS and KS
with values of the   parameters $\beta$  and $\lambda$  as given in Table 1.
\ref{tab:adjusted_parameters}.  
All other assumptions are those of Fig. \ref{fig:VRSEP_spc}. 
}{fig:VRSEP_cspc_fit}

\section{Conclusion}

Acoustic power
 and oscillation amplitudes of radial oscillations
computed for a solar model are compared
with solar seismic observations.
With the adopted approach for computing the acoustic power (paper I), 
we find that the entropy source term  significantly contributes to 
the excitation process of solar-like oscillations by turbulent convection,
in agreement with the results of GMK and simulations by  \citet{Stein91}. 

Our generalized formulation for the oscillations 
power $P_{\omega_0}$ injected into the modes allows us
 to investigate effects of changes in the turbulent components.

In the solar case, several kinetic energy spectra have been considered: 
the Kolmogorov spectrum (KS);
 the ``Raised Kolmogorov Spectrum'' (RKS) 
which is derived from observations of 
the solar granulation by  \citet{Roudier86};
the  ``Nesis Kolmogorov Spectrum'' (NKS) 
which is derived from observations performed by \citet{Nesis93}. 
We have, in addition, considered the ``Spiegel Spectrum'' (SS) 
which  takes into account the non-linear 
interaction between turbulent modes of low wavenumber
and the   ``Broad Kolmogorov Spectrum'' (BKS) 
which exhibits  
 intermediate properties between the SS and the NKS at low wavenumbers.

In a first step, the value of the free parameter $\beta$ is set to 
a value  loosely determined using  theoretical arguments 
and  $\lambda=1$ is assumed as in B92. 
The oscillation power spectra computed with 
the KS,  BKS,  NKS  and RKS significantly differ.
In contrast with Houdek's \citeyearpar{Houdek96} results,  only small
 differences between  the  KS and SS   are  found.
At high frequency, the RKS  gives rise to a frequency dependence 
of the velocity which more closely fits the 
frequency dependence of the solar observations  
than the other  spectra and in addition induces 
larger amplitudes. 
Moreover, the  RKS leads to a depth of 
excitation of the same order as the one found by  \citet{Stein00II}. 
This depth of excitation is found to be much
 smaller than the one associated with the KS.  
This result is agreement  with Osaki's \citeyearpar{Osaki93} 
statement that the source of excitation should be more 
concentrated than the extended source distribution found by B92. 

As a model for  the time spectrum of the turbulent eddies, 
we find that the gaussian form, which is usually considered,  
and the modified gaussian proposed by \citet{Musielak94} , both agree with the observations while the exponential form - 
also proposed by \citet{Musielak94} - is not satisfactory.

We have assumed an entropy spectrum which is expected  from the turbulent theory. 
This spectrum presents a steeper slope in the 
inertial-conductive region, i.e.
 above the wavenumber $k_c$.
 Only changes in the global amplitudes are observed when changing $k_c$:
 the amplitude increases with decreasing values of $k_c$. 
 Thus the value of $k_c$ is difficult to constrain because the global amplitude depends on the treatment of convection. Therefore  the   entropy spectrum expected from the theory cannot be  validated  with the solar seismic observations.
The value of $k_c$ has thus been set to
 the value suggested by the observations
 of the solar granulation by  \citet{Espagnet93}.

Assuming a theoretical value for  the free parameter 
$\beta$ (as suggested in Paper~I) 
for each spectrum, we find that the absolute amplitude 
remains much smaller than the observations 
(about $3-5$~cm/s). Because the 
formulation is very sensitive to  $\beta$, 
the observed differences in the  maximum amplitude  between 
the theoretical estimates and the observations 
are mainly related to the uncertainties 
in the determination  of $\beta$.	
Observations of solar granulation are not accurate enough to provide 
a very precise value of the parameter $\beta$.
 Also the product $\beta \lambda$ controls the depth of excitation
and therefore also the frequency dependence of the expected power.
Accordingly, $ \beta \lambda$ is given a value 
for each spectrum which best matches the frequency dependence of 
the power spectrum at low frequencies to the observations. 
We next determine the value of $\beta $ and $\lambda$ 
in order to adjust the maximum amplitude to the observations.
This in turn defines the eddy  time correlation and the scale at which energy
injection is initiated.

Doing so, all the turbulent spectra give rise to a  similar behavior 
for $P_{\omega_0}$ at low frequencies but the KS and NKS fit much better 
the observations   at high frequencies.  
The values of $\beta$ and $\lambda$ resulting from the 
fitting performed with the RKS, BKS and NKS  are found 
to be compatible with the  \citet{Nesis93} 
observations of the solar granulation. 
For the KS we find much larger values of $\beta$ and $\lambda$. 
The corresponding value of $\beta \lambda$ is not compatible 
with the estimated upper limit value of $\beta \lambda$. 
This suggests that the NKS is in better agreement with the observed 
surface velocity spectrum of the Sun

The agreement in using the NKS between theoretical results and solar observations  
is found to be much better than that obtained in previous work.
In particular at high frequencies, 
the power derived from the observations exhibits a slope of about  $-6$  above $\nu = 4.5$~mHz. 

GMK predicts a slope of about $-4.4$ and
  a slope of about $-1.0$ is found
 with B92's formulation. 
In the present work, we find a slope  $\log(P_{\omega_0})$ vs $\log(\nu)$ 
of about $-6$ in the frequency domain  $\nu \gtrsim 4.5$~mHz. 

It is not obvious that the above conclusions 
hold true for  other solar-like oscillating stars. 
In particular one expects $P_{\omega_0}$  to 
exhibit  different properties for stars with rotation and magnetic activity
which are different to that of the Sun and  the NKS might  
well not  not suitable to represent the actual turbulent spectrum in these stars.

Assuming 
the same properties of turbulence i.e. same values of the free parameters
  for a given turbulent spectrum, it is possible to apply our calibrated formulation
 to several potentially solar-like oscillating stars 
in order to investigate the  consequences of 
changing  the turbulent ingredients
 and to establish constrains  on  the stellar turbulent spectrum \citep{Samadi00b,Samadi00III}.

On the theoretical side, further tests and validation 
are needed. In particular, as suggested in Paper~I,
 3D simulations of the stellar convection 
will enable us to test some of the 
assumptions and approximations entering the present formulation \citep{Samadi00Phd}.
In addition, such simulations may enable us 
to constrain more accurately the  values of the free parameters 
than has been achieved here with  solar seismic data.

\begin{acknowledgements}
We thank F. Tran Minh and L. Leon for the use of the FILOU pulsation code and we are indebted to J.-P. Zahn for useful discussions. 
\end{acknowledgements}

\bibliography{../../biblio}
\bibliographystyle{apj}
\end{document}